\documentclass[iop]{emulateapj}

\usepackage{txfonts}
\usepackage{graphicx}

\usepackage{graphicx}

\usepackage{soul}

\usepackage{natbib}

\usepackage{natbib}

\begin{document}
\title{H$\alpha$ line profile asymmetries and the chromospheric flare velocity field}
\vskip1.0truecm
\author{
D.~Kuridze$^{1,5}$, M.~Mathioudakis$^{1}$, P.~J.~A.~Sim\~{o}es$^2$, L.~Rouppe van der Voort$^2$, 
M.~Carlsson$^3$, S. Jafarzadeh$^3$, J.~C.~Allred$^4$, A.~F.~Kowalski$^4$, M.~Kennedy$^1$, 
L.~Fletcher$^2$,   D.~Graham $^6$ \& F. P. Keenan$^1$} 
\affil
{$^1$Astrophysics Research Centre, School of Mathematics and Physics, Queen's University Belfast, BT7 1NN, Northern Ireland, UK.}
\affil{$^2$SUPA School of Physics and Astronomy, University of Glasgow, Glasgow G12 8QQ, U. K.}
\affil{$^3$Institute of Theoretical Astrophysics, University of Oslo, P.O. Box 1029 Blindern, N-0315 Oslo, Norway.}
\affil{$^4$ NASA/Goddard Space Flight Center, Code 671, Greenbelt, MD 20771.}
\affil{$^5$ Abastumani Astrophysical Observatory at Ilia State University, 3/5 Cholokashvili avenue, 0162 Tbilisi, Georgia,}
\affil{$^6$ INAF-Ossevatorio Astrofisico di Arcetri, I-50125 Firenze, Italy.}

\date{received / accepted }


\begin{abstract}

The asymmetries observed in the line profiles of solar flares can provide important diagnostics of the 
properties and dynamics of the flaring atmosphere. 
In this paper the evolution of the H$\alpha$ and Ca {\sc{ii}} 8542 {\AA}
lines are studied using high spatial, temporal and spectral resolution ground-based observations of an M1.1 flare
obtained with the Swedish 1-m Solar Telescope. The temporal evolution of the H$\alpha$ line profiles from the flare kernel 
shows excess emission in the red wing (red asymmetry) before flare maximum, 
and excess in the blue wing (blue asymmetry) after maximum. 
However, the Ca {\sc{ii}} 8542 {\AA} line does not follow the same pattern, showing only a weak red asymmetry during the flare.
RADYN simulations are used to synthesise spectral line profiles for the flaring atmosphere, and good agreement is found with the observations. 
We show that the 
red asymmetry observed in H$\alpha$ is not necessarily associated with plasma downflows, 
and the blue asymmetry may not be related to plasma upflows.
Indeed, we conclude that the steep velocity gradients in the flaring chromosphere modifies the wavelength of the central reversal in the H$\alpha$ line profile. 
The shift in the wavelength of maximum opacity to shorter and longer wavelengths generates the red and blue asymmetries, respectively.


\end{abstract}


\section{Introduction}

It is widely accepted that the vast majority of the solar flare radiative energy originates in the photosphere and chromosphere. 
The chromospheric radiation is dominated by hydrogen, calcium and magnesium lines, plus hydrogen continua, 
providing vital diagnostics on energy deposition rates in the lower atmosphere.  
One of the main characteristics of the flaring chromosphere is the centrally reversed H$\alpha$ emission line profile with asymmetric red and blue wings \citep[see the review by][]{ber07}.  
Earlier observations have shown that the red asymmetry in the H$\alpha$ line profile is more typical and is  
observed frequently during the impulsive phase of the flare \citep{sve76, tan83,ich84,wus89,fal97}. 
High resolution spectroscopy indicates that the red asymmetry  
dominates over the ribbons during the flare \citep{asa12,den13,hua14}.
However, blue asymmetries have also been reported and these occur mainly in the early stages of the flare \citep{sve62,sev68,hei94,mei97,yu06}.
Furthermore, blue and red asymmetries at different positions in the same flare ribbon, 
and the reversal of asymmetry at a same position in the ribbon have also been reported \citep{can90b,ji94,mei97}.
The Ca {\sc{ii}} 8542 {\AA} line profile, which goes into full emission during the flare, 
is also asymmetric \citep{li07}, but its asymmetric properties  
do not necessary coincide with those detected simultaneously in H$\alpha$ \citep{mei97}. 

Despite a number of  attempts to explain the observed asymmetries, their exact nature remains unclear. 
Theoretical line profiles based on static models of flare chromospheres do not show  
asymmetric signatures \citep{can84,fan93,che06}. This suggests that the asymmetries are related to the velocity field in the flaring atmosphere. 

\begin{figure*}[t]
\begin{center}
\includegraphics[width=17.9cm]{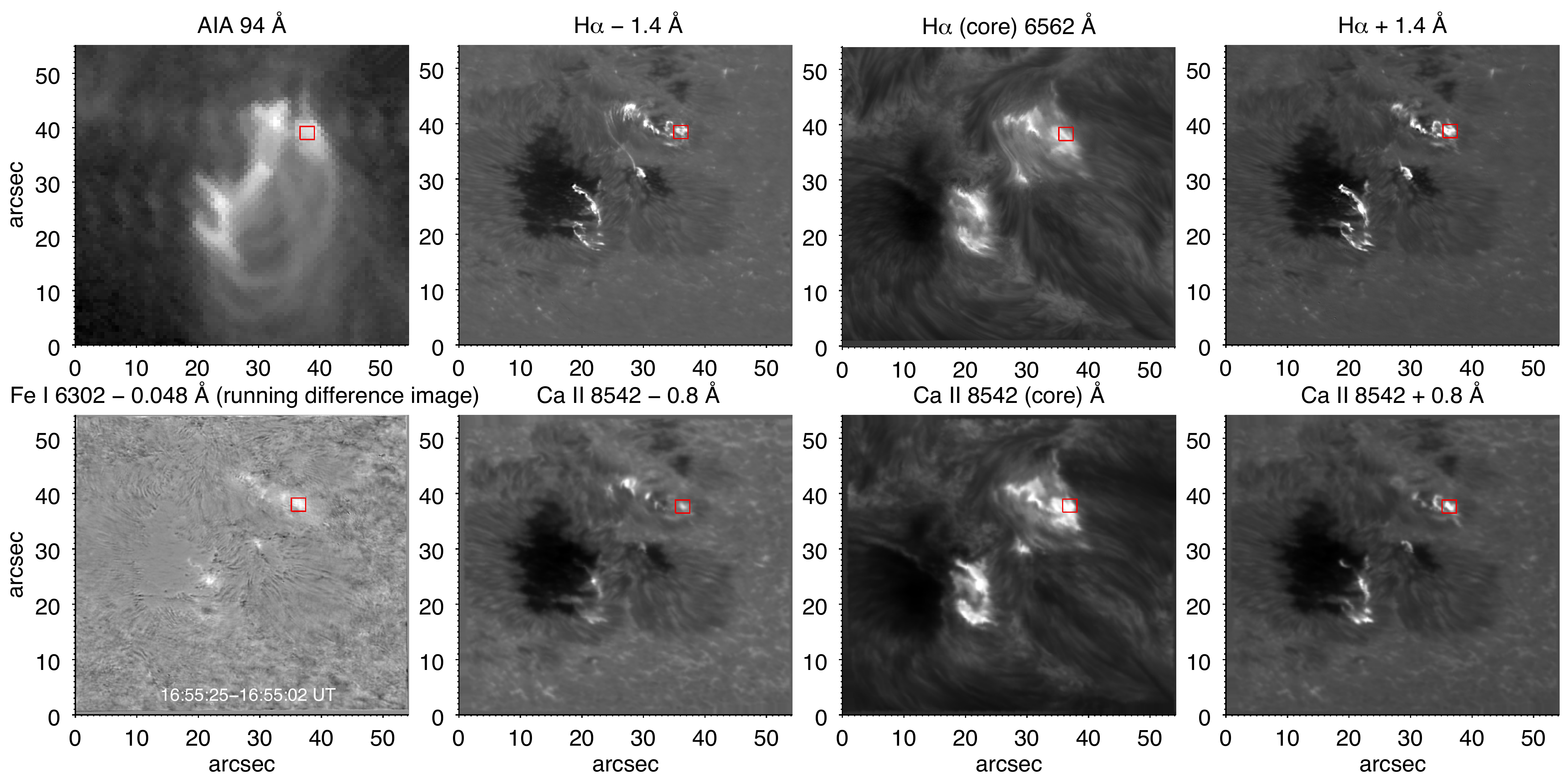}
\end{center}
\caption{H$\alpha$ and Ca {\sc{ii}} 8542 {\AA} wing and core images 
together with a Fe {\sc{i}} $\mathrm{6302-0.048~{\AA}}$ running difference image obtained with the CRISP instrument on the SST. 
A co-temporal SDO/AIA 94 {\AA} image of the 
flaring region is also shown. 
The red boxes indicate the flare kernel analysed in this paper co-spatial with the footpoint of the coronal loops observed in the AIA 94 $\AA$ image.}
\label{fig1}
\end{figure*}


The red asymmetries observed in chromospheric lines during flares are usually attributed to 
the downflow of cool plasmas in the flaring atmosphere, a process also known as   
chromospheric condensation \citep{ich84,wus89,bea92}. 
As the flare energy is deposited in the chromosphere, 
a strong evaporation of the heated plasma occurs \citep[e.g.][]{neu68,fis85,gra15}. Conservation of momentum requires the formation of a downflow in the form of chromospheric condensation \citep{ich84,can90a,mil09,hud11,ker15}. 
If the downflow occurs in the upper chromosphere, 
its effect would be a red-wing absorption which can lead to a blue asymmetry in the line profile \citep{gan93,hei94}. 
\cite{din96} have shown that both red and blue asymmetries in the H$\alpha$ line can be caused by downward moving plasma at different heights in the solar chromosphere. 
This blue asymmetry 
has also been interpreted in terms of chromospheric evaporation and filament activation \citep{can90b,hua14}. 


\cite{abb99} computed time-dependent H$\alpha$ and Ca {\sc{ii}} K line profiles with the radiative-hydrodynamic code 
\citep[RADYN;][]{car97} for weak (F10) and strong (F11) flare runs, 
and showed that asymmetries could be produced by the strong velocity gradients
generated during the flare. The velocity gradients create differences in the opacity between the red and blue wings of H$\alpha$, and the sign of the gradient 
determines whether the asymmetric emission appears to the blue or red side of the line profile.    
This process was first reported in the pioneering work of \cite{car97}, 
who studied the effects of acoustic shocks on the formation of the Ca {\sc{ii}} H line in a non-flaring atmosphere.
More recently, \cite{fat15} made a direct comparison between observations and RADYN simulations 
of an M3.0 flare. Their observed and synthesised H$\alpha$ and Ca {\sc{ii}} 8542 ${\AA}$ line profiles are asymmetric, especially during the early stages of the flare.   

In this paper we present high temporal, spatial and spectral resolution imaging spectroscopy of an M1.1 solar flare.  
We study the evolution of the H$\alpha$ and Ca {\sc{ii}} 8542 {\AA} line profiles of the flare kernel,
and compare these with synthesised line profiles obtained with a radiative hydrodynamic simulation.   
The line contribution functions and the velocity field in the simulated atmosphere allow us to disentangle the nature of the observed line asymmetries. 


\section{Observations and data reduction}
\label{sect:setup}

The observations presented in this paper were undertaken between 16:27 and 17:27 UT on 2014 September 6 with the CRisp Imaging
SpectroPolarimeter \citep[CRISP;][]{shr06,shr08} instrument, mounted on the Swedish 1 m Solar Telescope \citep[SST;][]{shr03a}
on La Palma.  Adaptive optics were used throughout the observations, consisting of a tip-tilt mirror 
and a 85-electrode deformable mirror setup that is an upgrade of the system described in \cite{shr03b}. 
The observations comprised of spectral imaging in the H$\alpha$~6563~{\AA}, Ca {\sc{ii}} 8542 ${\AA}$ lines and  Stokes V maps in Fe {\sc{i}} $\mathrm{6302~at- 0.048~{\AA}}$.
All data were reconstructed with Multi-Object Multi-Frame Blind Deconvolution \citep[MOMFBD;][]{noo05}. 
We applied the CRISP data reduction pipeline as described in \cite{cru15}.
Our spatial sampling is 0$''$.057 pixel$^{-1}$
and the spatial resolution is close to the diffraction limit of the telescope for many images in the time series.
The H$\alpha$ line scan consists of 15 positions symmetrically sampled with 0.2 ${\AA}$ steps from line core, with the 
Ca {\sc{ii}} 8542 ${\AA}$  scan consisting of 25 line positions with 0.1 ${\AA}$ steps.
A full spectral scan had an acquisition time of a 11.54 s, which is the temporal cadence of the time series.
We made use of CRISPEX \citep{vis12}, a versatile widget-based tool for effective viewing and 
exploration of multi-dimensional imaging spectroscopy data.

The flare was also observed by the Reuven Ramaty High Energy Solar Spectroscopic Imager \citep[RHESSI;][]{lin02} 
and Fermi/GBM \citep{mee09}. In Section 4 we describe the analysis of RHESSI data, noting that similar results were found from Fermi/GBM. 
The HXR spectral analysis was performed using OSPEX \citep{sch02} to estimate the power $P_\mathrm{nth}$ deposited 
in the chromosphere by the non-thermal electrons, assuming a thick-target model. 


\begin{figure}[t]
\begin{center}
\includegraphics[width=8.67 cm]{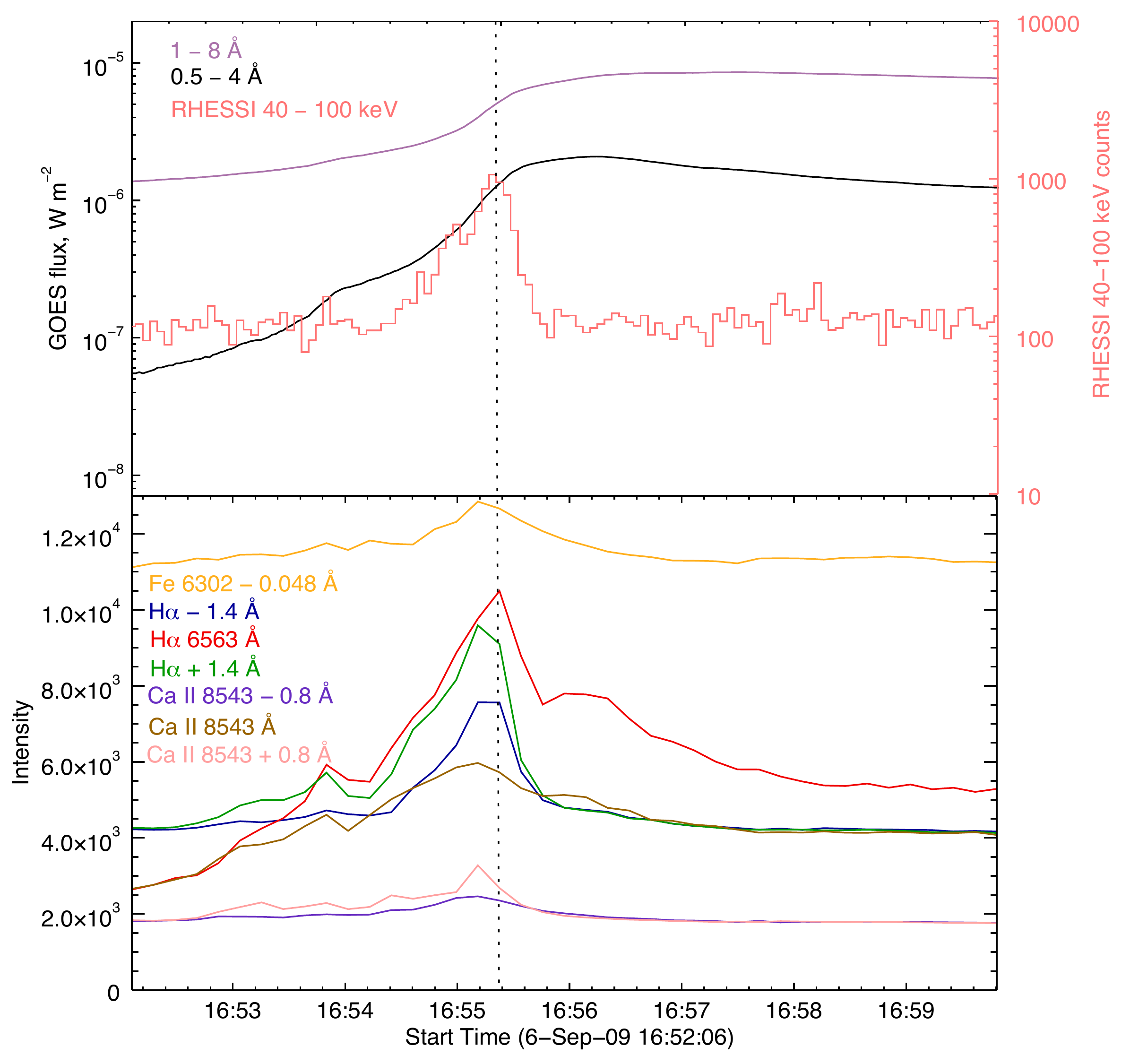}
\caption{Top panel: GOES X-ray and RHESSI $\mathrm{40-100~keV}$ hard X-ray lightcurves. Bottom panel: Lightcurves of the flaring region marked with the red boxes in Figure~\ref{fig1}.}
\label{fig2}
\end{center}
\end{figure}

\begin{figure*}[t]
\begin{center}
\includegraphics[width=16.2 cm]{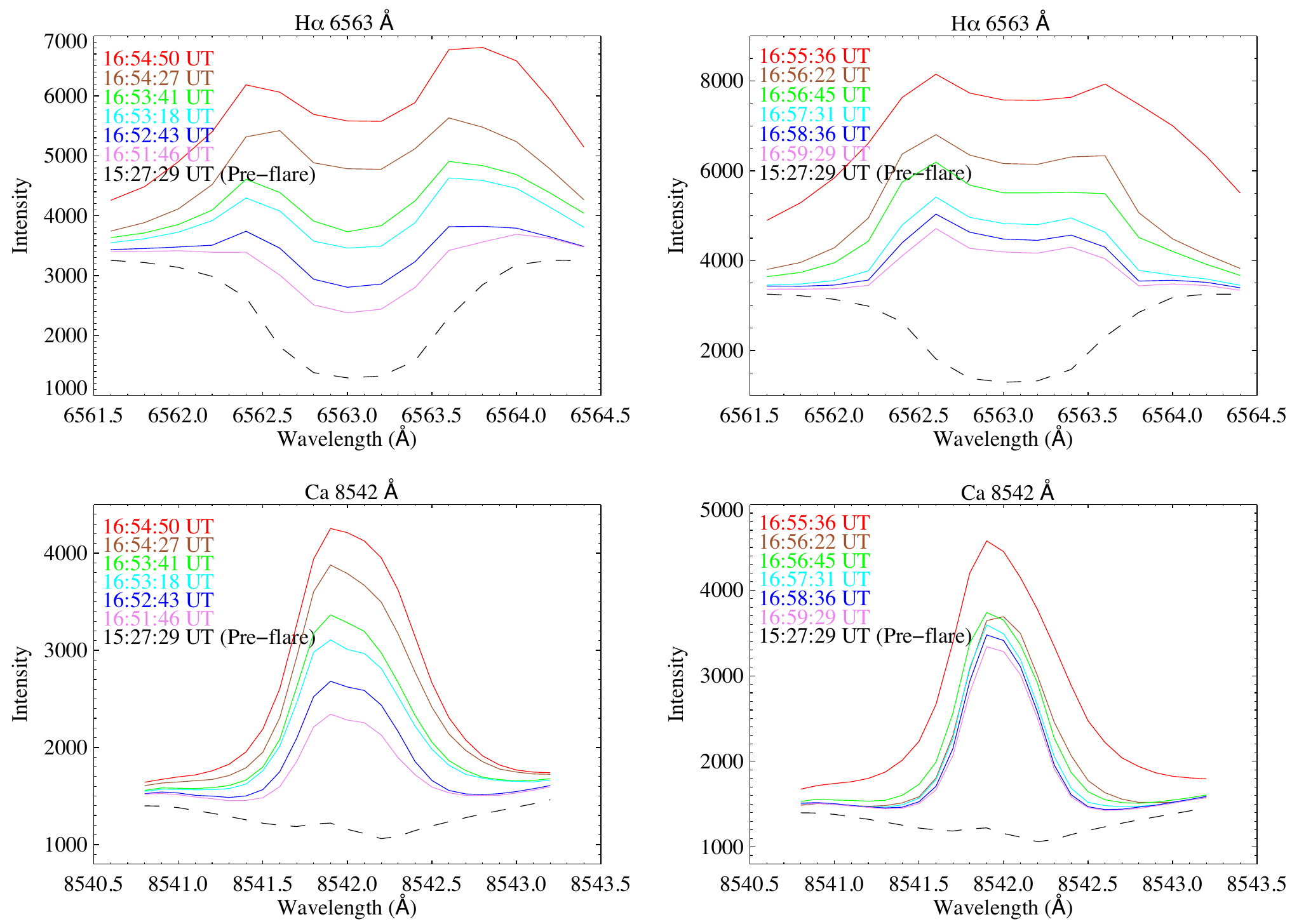}
\caption{The temporal evolution of the H$\alpha$ and Ca {\sc{ii}} 8542 {\AA} line profiles before the flare maximum (left panels) and after the flare maximum (right panels) for the region outlined with the red boxes in Figure~\ref{fig1}.}
\label{fig3}
\end{center}
\end{figure*}

\begin{figure}[t]
\includegraphics[width=8.6 cm]{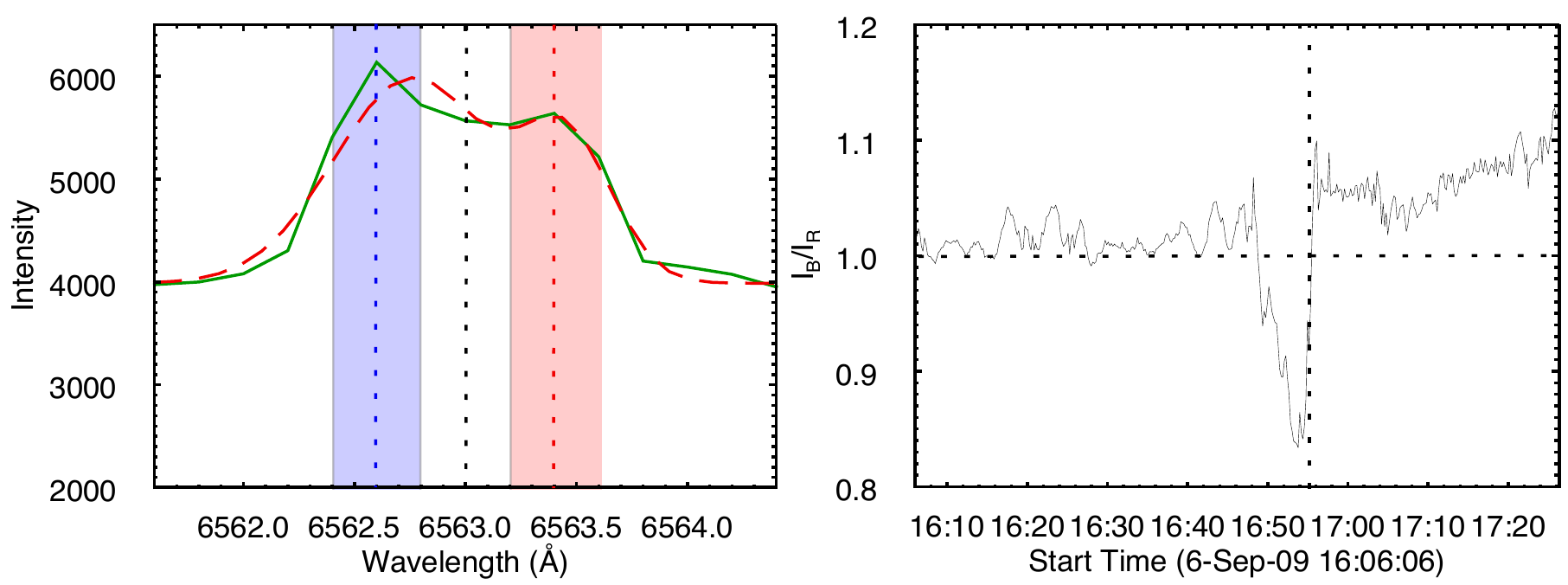}
\caption{The left panel shows the same H$\alpha$ line profile presented in the top left panel of Figure~\ref{fig3} with green line. 
The dashed red line shows the double-Gaussian fit. 
The right panel shows the evolution of line profile asymmetry, $I_B/I_R$, calculated by dividing the intensities of the blue and red peak centroid positions 
integrated over the $\pm0.2~{\AA}$ wavelength range (red and blue areas shown on the left panel).
Values of $I_B/I_R$ are normalised by the mean intensity ratios of the non-flaring atmospheric profile, 
such that $I_B/I_R\approx 1$ corresponds to no asymmetries in the H$\alpha$ spectra.}
\label{fig4}
\end{figure}


\section{H$\alpha$ \& Ca {\sc{ii}} 8542 \AA\ Line Asymmetries}

The two ribbon M1.1 flare was observed in active region NOAA 12157 located 
at heliocentric coordinates $\sim(-732'', -302''$).  
Figure~\ref{fig1} shows the flare images in H$\alpha$ and Ca {\sc{ii}} 8542 ${\AA}$ line core and wing positions, 
together with a photospheric Fe {\sc{i}} $\mathrm{6302-0.048~{\AA}}$ running difference image. 
An examination of the temporal evolution of the flare ribbons suggests that they remain stationary in the image plane during the flare. 
We identified a bright kernel in the CRISP images, highlighted with the red boxes in Figure~\ref{fig1}, 
which is co-spatial with the footpoint of the coronal loops observed in AIA 94 $\AA$ images (Figure~\ref{fig1}). 
Lightcurves of the region show that the emission from the H$\alpha$ and Ca {\sc{ii}} 8542 ${\AA}$ line wings and the photospheric Fe {\sc{i}} 
line between 16:54 and 16:56 UT increases, peaking at 16:55:24 (Figure~\ref{fig2}).  
We find no evidence for time delays between the different wavelengths in excess of the 11.54 s temporal resolution element.  

In Figure~\ref{fig3} we show the H$\alpha$ and Ca {\sc{ii}} 8542 ${\AA}$ line profiles averaged over the $\sim$2$''$$\mathrm{\times}$2$''$ area marked with the red boxes in Figure~\ref{fig1}. 
Shortly after the flare onset the flaring chromosphere produces a  wide H$\alpha$ emission profile with a central reversal.  
The temporal evolution of the centrally reversed H$\alpha$ profiles shows excess emission in the red wing 
(red asymmetry) with nearly unshifted line center before the flare maximum (top left panel of Figure~\ref{fig3}). 
However, after the flare maximum the blue peak becomes higher than the red peak and
hence a blue asymmetry is formed (top right panel of Figure~\ref{fig3}).
Furthermore, the line centers of these blue asymmetric H$\alpha$ profiles are shifted to longer wavelengths (top right panel of Figure~\ref{fig3} and Figure~\ref{fig4}).

To quantify the asymmetry of the H$\alpha$ line profiles we use a technique similar to that described in \cite{mei97} and \cite{dep09}. 
We identify the central wavelength of the red and blue peaks   
and compute the intensities in these positions integrated over a $\pm0.2$ {\AA} interval (left panel of Figure~\ref{fig4}). 
These intensities are given as 
\begin{equation}
I_B=\sum^{\lambda_{0B}+\delta\lambda}_{\lambda_{0B}-\delta\lambda}I_{\lambda},~~~~I_R= \sum^{\lambda_{0R}+\delta\lambda}_{\lambda_{0R}-\delta\lambda}I_{\lambda},
\label{eq2}
\end{equation}
where $\lambda_{0B}$ and $\lambda_{0R}$ are the centroids of the blue and red peaks, respectively,  
and $\delta\lambda$ the wavelength range over which the intensities are integrated.  
The ratio $I_B/I_R$ allows us to study the evolution of the asymmetries as a function of time (right panel of Figure~\ref{fig4}), and is  
normalised to the mean intensity ratios of the non-flaring atmospheric profile computed over 
the same wavelength range, such that $I_B/I_R\approx 1$ correspond to no asymmetries in the H$\alpha$ line profile. 
Figure~\ref{fig4} (right panel) shows that at around $\sim$16:50 UT a strong red asymmetry begins to appear, increasing to the flare maximum ($\sim$16:55 UT).
The red asymmetry then decreases very quickly (within about 2 minutes) with the blue asymmetry appearing and maintained until the end of the observations.    

The Ca {\sc{ii}} 8542 {\AA} line has a slightly redshifted emission profile of about 0.1~{\AA} 
(corresponding to a velocity $\sim\mathrm{3.5~km~s^{-1}}$), and in contrast to H$\alpha$ 
does not reveal the asymmetry reversal during the flare (bottom panels of Figure~\ref{fig3}). 


\section{Hard X-rays}
\label{HXR}

Hard X-ray emission is a typical signature for the presence of accelerated electrons in flares. 
Assuming that the electron distribution 
$F(E)$ has a power-law form $AE^{-\delta}$ (electrons s$^{-1}$ keV$^{-1}$), where $A$ is the normalisation factor (proportional to the total electron rate), 
$E_C$ is the low energy cutoff, and $\delta$ is the spectral index, a lower limit for the total power contained in the distribution can be estimated from: 

\begin{equation}
P_\mathrm{nth}(E \ge E_C) = \int_{E_C}^\infty E F(E) dE \ \mathrm{erg s^{-1}}
\label{eq3}
\end{equation}

\begin{figure}
\begin{center}
\includegraphics[width=8cm]{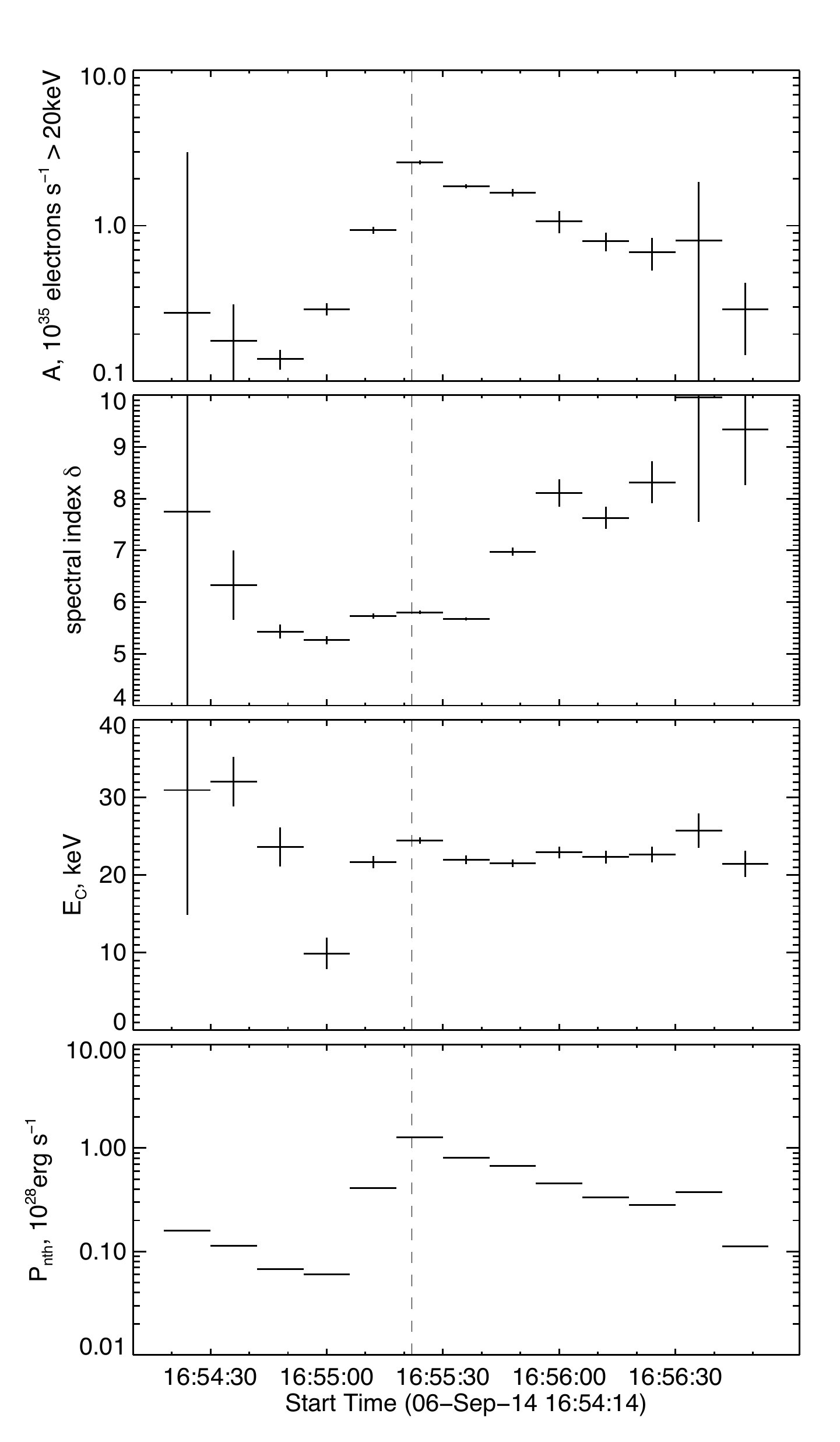}
\end{center}
\caption{Fitting results of RHESSI HXR data: normalisation factor $A$, spectral index $\delta$, low energy cutoff $E_C$ and the total power contained 
in the electron distribution $P_\mathrm{nth}$, given by Eq. \ref{eq3}. The vertical dashed line indicates the time of the maximum HXR emission.}
\label{fig5}
\end{figure}

\begin{figure}
\begin{center}
\includegraphics[width=8.3cm]{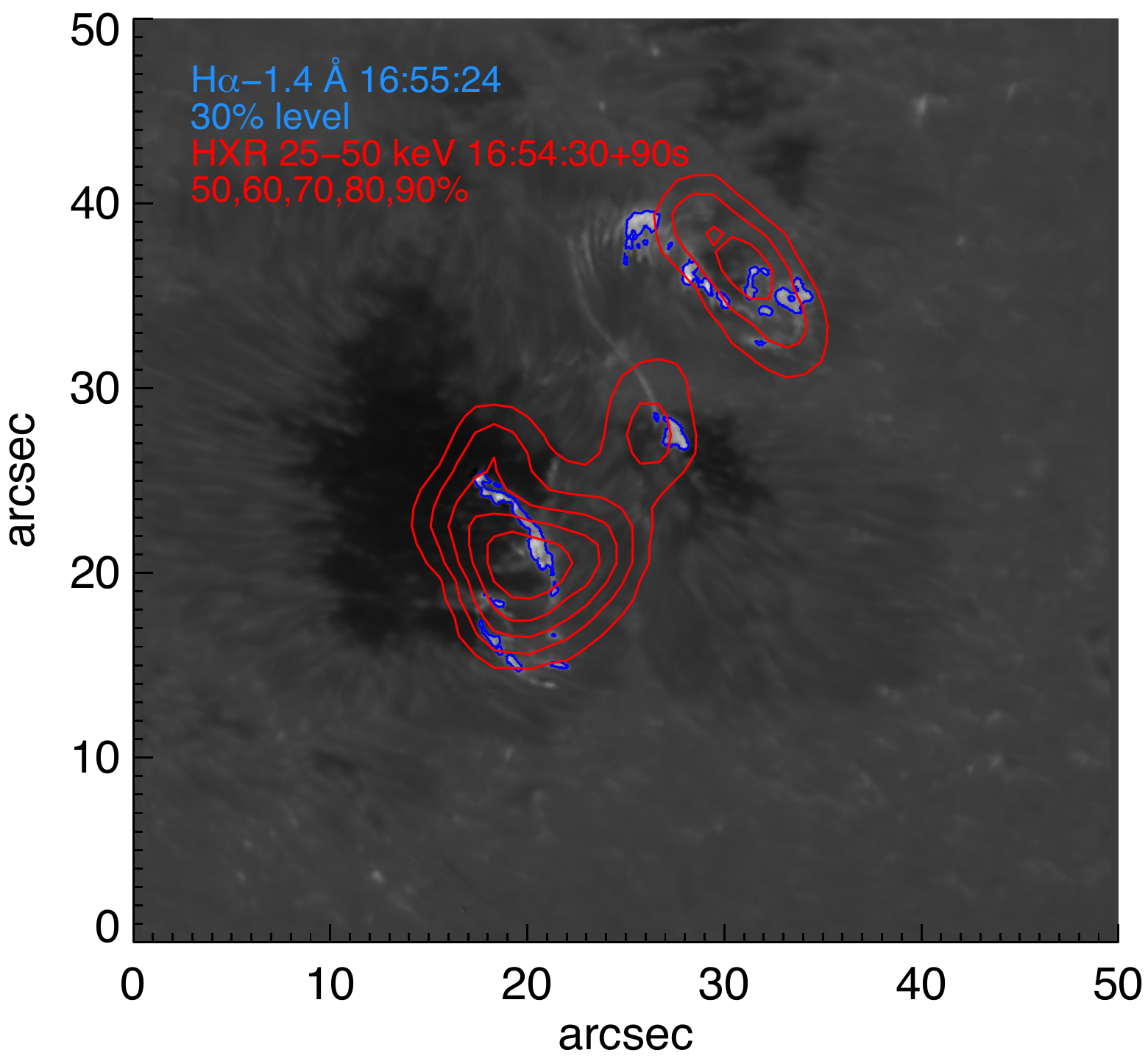}
\end{center}
\caption{H$\alpha$ - 1.4~\AA ~image at the flare peak (16:55:24~UT), with blue contours showing the 30\% level of the image maximum, 
overlaid with a RHESSI 25-50 keV reconstructed image, shown as red contours at 50, 60, 70, 80, 90\% of the maximum.} 
\label{fig6}
\end{figure}

We fitted the HXR spectrum obtained by RHESSI with a isothermal plus thick-target model (as described above, using the OSPEX function {\tt \verb$thick2_vnorm$}), 
and calculated $P_\mathrm{nth}$ throughout the impulsive phase, integrating the counts in bins of 12 seconds. 
RHESSI front detectors 1, 3, 6, 8 and 9 were employed, and the fitting results for $A$, $\delta$, $E_C$ and $P_\mathrm{nth}$ are shown in Figure~\ref{fig5}. 
We found $P_\mathrm{nth} \approx 1.3\times10^{28}$ erg s$^{-1}$ at the time of maximum of HXR emission, between 16:55:18 and 16:55:30~UT. 
To obtain the energy flux (erg s$^{-1}$ cm$^{-2}$) deposited into the chromospheric source, one divides the power of the non-thermal electrons 
$P_\mathrm{nth}$ by the footpoint area. The reconstructed images for the HXR emission (using CLEAN, \cite{hur02}, detectors 3--8, beam width factor of 1.5) 
show three main HXR sources in good association with the main locations of the optical and UV emission, as can be seen in Figure~\ref{fig6}. 
However, the HXR sources are unresolved and the area obtained from the RHESSI images would be an overestimate of the actual sizes of those regions. 
The area of the ribbons was hence determined by contouring a region with intensity above 30\% of the maximum of the H$\alpha$-1.4~\AA\ image near the HXR peak (Figure~\ref{fig5}). 
We obtained an area of $0.7 \times 10^{17}$ cm$^2$. 
Thus, at the time of the maximum HXR emission, which also coincides with the maximum of the emission in H$\alpha$, Ca {\sc ii}, Fe {\sc i} 
(see Figure~\ref{fig2}), the average energy flux delivered by the electrons into the chromosphere is around $1.9\times 10^{11}$ erg s$^{-1}$ cm$^{-2}$.

\section{Simulated Line Profiles}

The short duration of photospheric enhancements (Figure~\ref{fig2}) 
and the stationary nature of the flare ribbon makes this event a good candidate for a comparison with a dynamic model.  

We use the radiative-hydrodynamic code RADYN to compute H$\alpha$ and Ca {\sc{ii}} 8542 \AA\ line profiles \citep{car97,all05}. 
The simulation was performed for a strong beam with $\mathrm{F}=10^{11}\mathrm{ergs~cm^{-2}~s^{-1}}$ (also known as F11 flare) and an isotropic pitch angle distribution in the forward hemisphere
with the Fokker-Planck solution to the nonthermal electron distribution \citep{all15}.
A constant heating flux was applied for 20 sec, and the atmosphere was allowed to relax for 40 additional seconds. 
We used a power-law index  and a low energy cut-off of  $\delta$=4.2 and $E_c$ = 25 keV, respectively.  
The beam parameters used in the model are very close to the ones estimated from the HXR data (Section~\ref{HXR}).

\begin{figure*}
\includegraphics[width=18.0 cm]{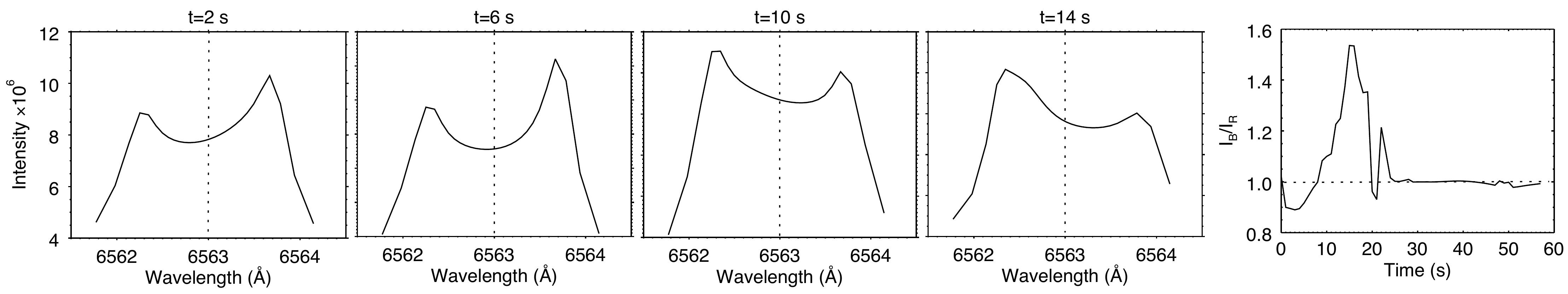}
\caption{The temporal evolution of the synthesised H$\alpha$ profiles during an F11 flare. The vertical dotted lines indicate the line center. 
The right panel shows the evolution of line profile asymmetry, $I_B/I_R$, calculated with the method described in Section 3.}  
\label{fig7}
\end{figure*}

Figure~\ref{fig7} shows the temporal evolution of the synthesised H$\alpha$ profiles from the F11 flare model,   
which indicates a behaviour similar to that in the observations. 
Increased emission in the red side of the profile (i.e. red asymmetry) occurs when the H$\alpha$ line centre is shifted to the blue, 
whereas during the blue asymmetry the line centre is shifted to the red.
The right panel of Figure~\ref{fig7} shows the evolution of the line profile asymmetry, $I_B/I_R$, calculated using the method described in Section~3. 
Simulated Ca {\sc{ii}} 8542 {\AA} line profiles are in emission during the flare with an extended blue wing and a slightly blueshifted center  
(Figure~\ref{fig10}).

To understand the formation of the asymmetric line profiles, we need to examine the line contribution functions \citep{car97}. 
These are the intensities emitted in specific wavelengths as a function of height. 
\cite{car97} introduced a formal solution of the radiative transfer equation for the emergent intensity in terms of the contribution function, $C_{I}$, as 
\begin{equation}
I_{\nu}=\int_{z_0}^{z}C_{I}dz=\int_{z_0}^{z}S_{\nu}\tau_{\nu}e^{-\tau_{\nu}}\chi_{\nu}/\tau_{\nu}dz,
\label{eq1}
\end{equation}
where $z$ is the atmospheric height, and $S_{\nu}$, $\tau_{\nu}$ and $\chi_{\nu}$ are the source function, 
optical depth and opacity (linear extinction coefficient), respectively. 
Traditionally, this function is attuned to the three frequency ($\nu$) dependent physically meaningful terms:  $S_{\nu}$, $\tau_{\nu}e^{-\tau_{\nu}}$ and $\chi_{\nu}/\tau_{\nu}$,
where $S_{\nu}$ is defined as the ratio of the emissivity over the opacity of the atmosphere. 
The $\tau_{\nu}e^{-\tau_{\nu}}$ term expresses an attenuation caused by optical depth and peaks near $\tau=1$,
while the ratio $\chi_{\nu}/\tau_{\nu}$ is large where there are many emitting particles at low optical depths.

\begin{figure*}
\begin{center}
\includegraphics[width=13.5cm]{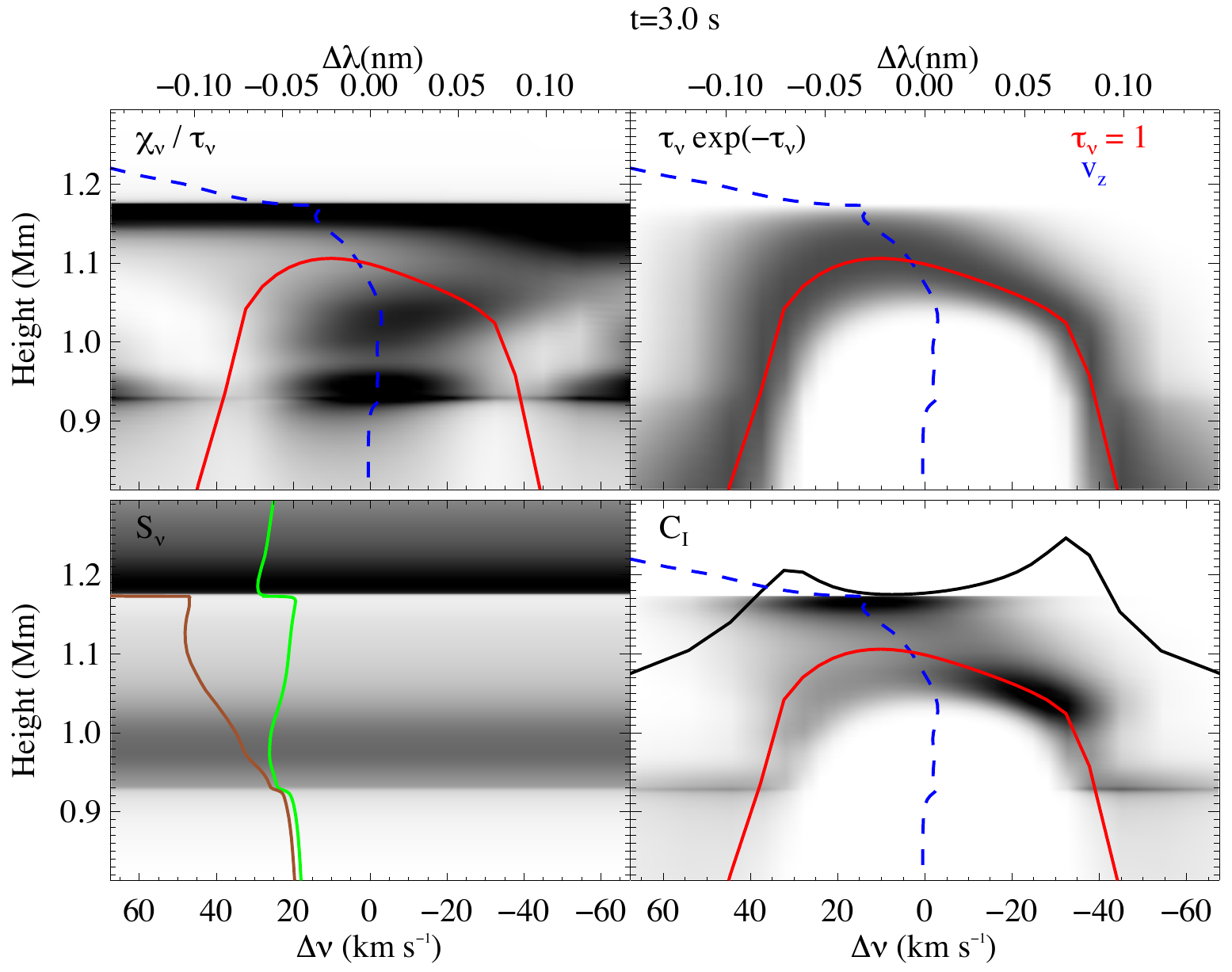}
\end{center}
\caption{Components of the intensity contribution function for the H$\alpha$ line together with the contribution function itself 
after 3 s of F11 flare heating. The diagrams are plotted in inverse gray scale so that darker shades indicate higher intensities. 
The line profile is overplotted in the contribution function diagram (bottom right panel) as a black line. 
Red lines indicate the height at which $\tau=1$. The vertical velocity structure of the plasma is overplotted as blue dashed lines. 
Positive velocity correspond to plasma upflows.
The source function (green line) and Planck function (brown line) at the line center are overplotted in the bottom left panels.}
\label{fig8}
\end{figure*}

\begin{figure*}
\begin{center}
\includegraphics[width=13.5cm]{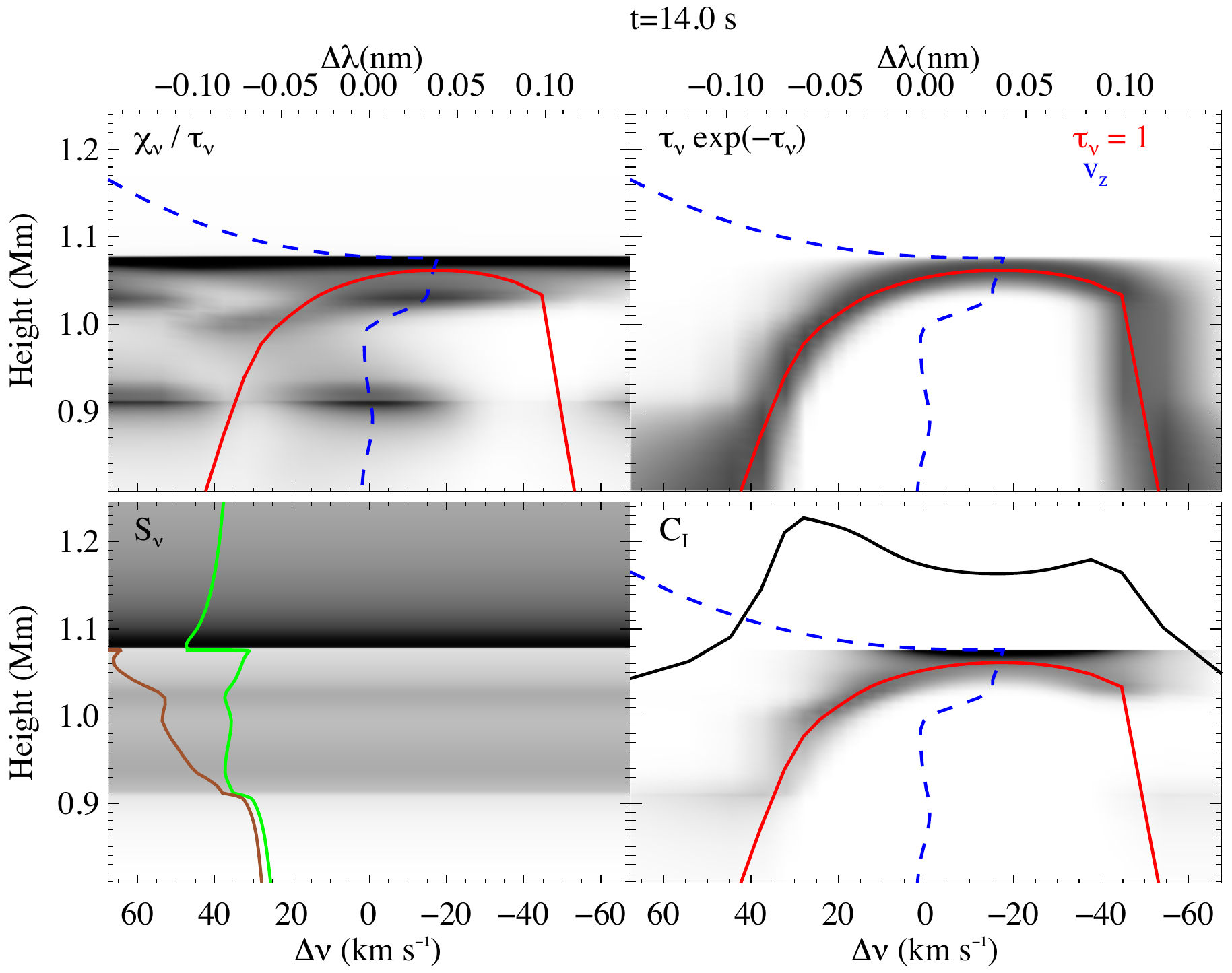}
\end{center}
\caption{Components of the intensity contribution function together with the contribution function itself for the H$\alpha$ line 
after 14 s of F11 flare heating. The line styles and notations are the same as in Figure~\ref{fig6}.}
\label{fig9}
\end{figure*}

In Figures~\ref{fig8},~\ref{fig9},~and~\ref{fig10} we present line contribution functions for H$\alpha$ and Ca {\sc{ii}} 8542 {\AA} 
using the classic diagrams introduced by \cite{car97}.
The diagrams are plotted in inverse gray scale with darker shades showing higher intensities. 
Line profiles are shown as dark lines in the bottom right panels. 
Red lines in the diagrams show the height at $\tau=1$ while the vertical velocity structure of the plasma is plotted as a blue dashed line. 
We note that positive values for velocities correspond to plasma upflows. 
The source functions (green lines) and Planck functions (brown lines) at line center are overplotted 
in the bottom left panels to show the height of non-LTE decoupling in the atmosphere.

Figure~\ref{fig8} contains a snapshot of the parameters that contribute to the H$\alpha$ line profile and the formation of the red asymmetry 3 s into the F11 flare run.  
It shows that the line wings and line core are formed at $\mathrm{0.9-1.0}$ and $\mathrm{1.1-1.18}$ Mm, respectively.   
The line source function has a local maximum near the wing formation height, and decreases towards 
the upper layers where the line core is formed. 
This leads to increased emission in the line wings, relative to the line core, hence the centrally reversed line profiles. 
The velocity structure of the flaring atmosphere shows that the region of H$\alpha$ wing formation has a weak downflow velocity of around $\mathrm{1-3~km~s^{-1}}$. 
However, the region of core formation has a moderately 
strong upflow with a peak value of $\mathrm{\sim15~km~s^{-1}}$ at 1.17~Mm (Figure~\ref{fig8}).
This upflow shifts the line core of the centrally reversed H$\alpha$ profile to the blue (bottom right panel).  
Therefore, the central wavelength of maximum opacity is also shifted to the blue.
Consequently, the $\tau=1$ layer is formed higher and the $\chi_{\nu}/\tau_{\nu}$ is smaller in the blue wing, and 
hence blue wing photons are absorbed more easily compared to their red wing counterparts.    

Figure~\ref{fig9} shows similar diagrams 14 s into the F11 flare run. Now the H$\alpha$ line core is formed deeper 
into the atmosphere, indicating that the chromosphere is compressed by the transition region which 
moves downward from 1.18 Mm to 1.08 Mm over about 10 s (see Figures~\ref{fig8},~\ref{fig9}).
At the core formation height (at around 1.08 Mm) the atmosphere has developed a moderately strong downflow (condensation) 
of about $\mathrm{-18~km~s^{-1}}$ (downward velocities are negative) with a negative velocity gradient.
This shifts the central reversal (maximum opacity) to the red side (Figure~\ref{fig9}).  
The higher lying plasma of the line core absorbs red photons, increasing the opacity in the red wing. 
Therefore, the red wing photons produced between 0.9-1.05 Mm can no longer escape as 
readily as at t=3~s because the $\tau=1$ curve has shifted to the red due to the downflowing material above.
As a result, the $\tau=1$ curve is formed higher in the atmosphere at the red wing and $\chi_{\nu}/\tau_{\nu}$ 
is larger at the blue side of the core.  
This makes the contribution function, and hence the emission, larger in 
the blue wing and the blue asymmetry is established (bottom right panel of Figure~\ref{fig9}).

\begin{figure}[t]
\begin{center}
\includegraphics[width=8.7cm]{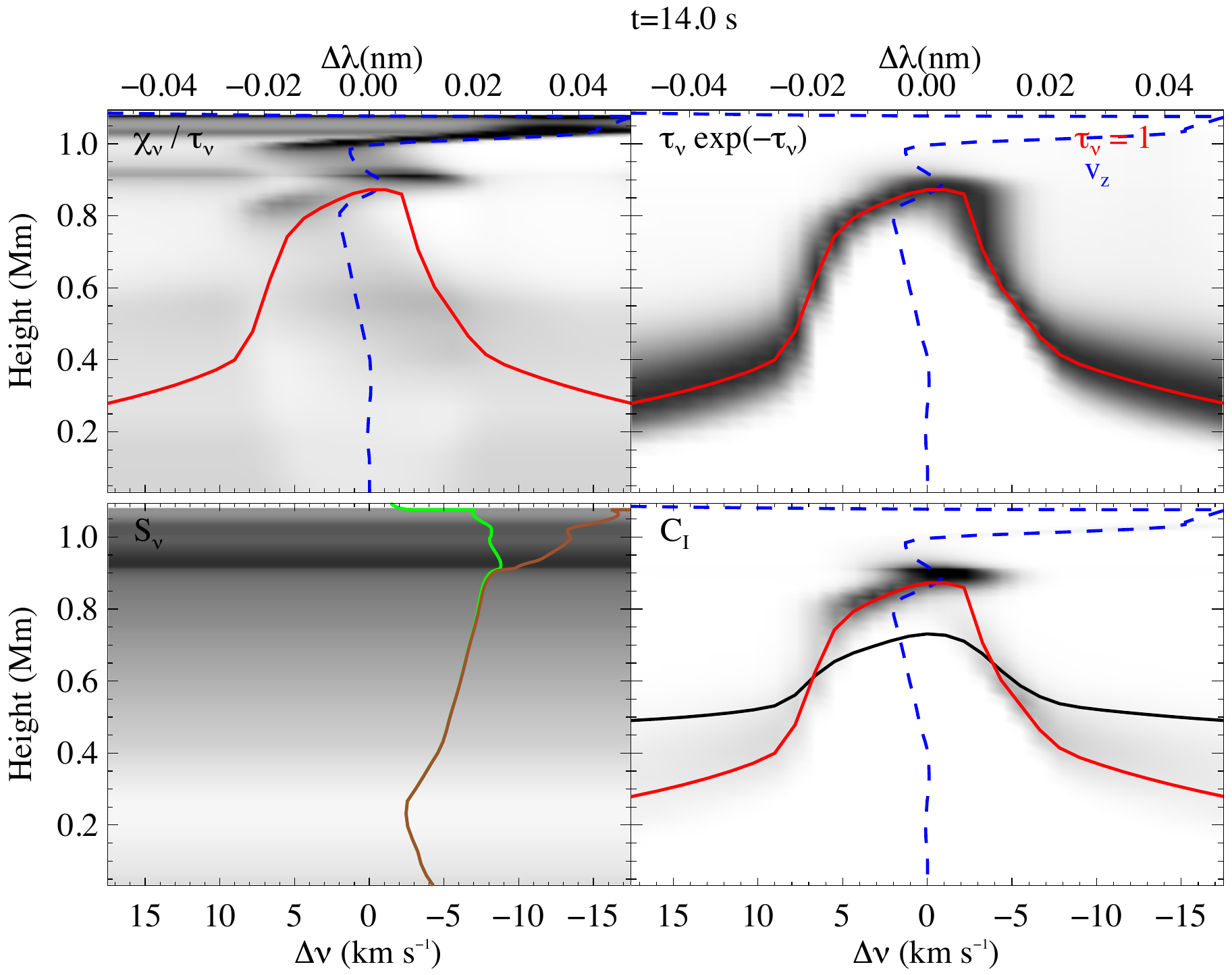}
\end{center}
\caption{Components of the intensity contribution function for the Ca {\sc{ii}} 8542 {\AA} line together with the contribution function itself 
after 14 s of F11 flare heating. The line styles and notations are the same as in Figure~\ref{fig6}.}
\label{fig10}
\end{figure}
\vspace{3mm} 

Figure~\ref{fig10} shows a similar diagram for the Ca {\sc{ii}} 8542 {\AA} line 14 s into the F11 flare. 
This line is much narrower and formed deeper into the atmosphere (below 1~Mm) 
than H$\alpha$.  The velocity field along this height is very weak ($\mathrm{\sim\pm1-2~km~s^{-1}}$) during the flare,
and the line source function has a maximum at the height of core formation ($\mathrm{\sim0.9~Mm}$) 
(bottom left panel), 
so the higher emission originates from the line core. 
An upflow velocity field below $\mathrm{\sim0.8~Mm}$ produces a slightly extended blue wing and creates a blue asymmetry.
However, a weak downflow at the height of core formation ($\mathrm{\sim0.9~Mm}$) shifts the line center to the red.


\section{Discussion and conclusion}

The temporal evolution of the H$\alpha$ line profile of the flare kernel 
revealed excess emission in the red wing with unshifted line core before flare maximum, 
and excess emission in the blue wing with a red shifted line core after flare maximum. 
Numerical simulations show that in an atmosphere without velocity fields, the centrally reversed H$\alpha$ profile is symmetric with respect to the line core \citep{can84,fan93,che06}.
However, the dynamic models take into account the mass motions associated with the processes of evaporation and condensation of the flaring material, 
which generate strong asymmetric signatures.


We emphasize that the analysis presented here has shown that the red asymmetry observed in the H$\alpha$ 
line profile is not necessarily associated with plasma downflows, and the blue asymmetry may not be related to plasma upflows. 
Thus, using these line asymmetries as a direct measure of Doppler velocities can lead to inaccurate results.

Motivated by the close match between 
simulations and observations, 
we analysed components of intensity contribution functions for the H$\alpha$ and Ca {\sc{ii}} 8542 ${\AA}$ 
lines \citep[based on the approach presented in][]{car97,abb99}  
We find that in the early stages of the flare the line core of the centrally reversed H$\alpha$ profile is blueshifted  
due to the strong upflows at the height of core formation (bottom right panel of Figure~\ref{fig8}).
The velocity decreases downward toward the 
wing formation regions, producing a steep positive velocity gradient.  
This gradient modifies the optical depth of the atmosphere in a way that higher lying (core) atoms 
absorb photons with shorter wavelengths (blue wing photons) and the red asymmetry is formed (Figure~\ref{fig8}).  
In the later stages the velocity field becomes dominated by the downflows due to the plasma condensation with strong negative velocity gradients (Figure~\ref{fig9}). 
This shifts the absorption peak (wavelength of maximum opacity) to the red  (bottom right panel of Figure~\ref{fig9}), 
so the higher lying core plasma now absorbs red photons and produces a blue asymmetry.


The Ca {\sc{ii}} 8542 {\AA} line in the F11 model is in emission and formed deeper into the atmosphere
than H$\alpha$.  The weak velocity field along its formation height 
produces an extended blue wing in the line profile (bottom right panel of Figure~\ref{fig10}), in agreement with the observations. 
The simulated profile shows a weak downflow at the height of core formation which shifts the line center to the red (bottom right panel of Figure~\ref{fig10}). This is not detected in the observed spectra. 

The main difference between the simulations and the observations presented in this work 
is that the core of the observed H$\alpha$ profiles remains unshifted when the red asymmetry is detected (top left panel of Figure~\ref{fig3}). 
This may be due to the presence of a much more complex velocity field with several unresolved condensation/evaporation patterns, 
which can produce an effectively unshifted or unresolved line core. 
The simulations show that when the red asymmetry is formed, there are weak downflows in the region between the core and 
wing formation heights (see velocity field in Figure~\ref{fig8}). However, these downflows are not strong enough to have a significant effect on the line profiles.  

It should be noted that the red asymmetry in H$\alpha$ begins at 16:50 (right panel of Figure~\ref{fig4}), whereas RHESSI hard X-ray 
emission starts to increase after 16:54 UT (top panel of Figure~\ref{fig2}). 
This suggest that the H$\alpha$ is responding to pre-impulsive phase heating. 
In the pre-impulsive phase flare reconnection can produce direct heating instead of non-thermal particle heating.
That could drive gentle, rather then explosive, chromospheric evaporation
which can produce an enhanced red-wing emission in a manner similar to the one presented in Figure~\ref{fig8}.
However, the difference between simulations and observations could be related to that the red asymmetry in simulations is a result of explosive evaporation produced by non-thermal particle heating, 
whereas in the observations the red asymmetry is suggested to be produced with a more gentle evaporation at the pre-impulsive phase heating.

Furthermore, the evolution timescale of the observed H$\alpha$ line profile is longer ($\mathrm{\sim5~min}$)
than that of the simulated line profile ($\mathrm{\sim20~s}$).
Additional direct comparisons of the improved spectral resolution observations and more realistic simulations 
are necessary to understand the formation of spectral line asymmetries in the flaring solar chromosphere. 


\begin{acknowledgements}
The research leading to these results has received funding from the European Community's Seventh Framework Programme (FP7/2007-2013) under grant agreement no. 606862 (F-CHROMA).
The Swedish 1-m Solar Telescope is operated on the island of La Palma by the Institute for Solar Physics (ISP) of Stockholm University in the Spanish Observatorio del Roque de los Muchachos of the Instituto de Astrof\'isica de Canarias.
\end{acknowledgements}

\end{document}